\documentclass[%
 reprint,
 superscriptaddress,
 amsmath,amssymb,
 aps,
]{revtex4-1}

\usepackage{graphicx}
\usepackage{bm}
\usepackage{braket}
\usepackage{comment}
\usepackage{amsthm}
\usepackage{multirow}
\usepackage{hyperref}

\newcommand{\average}[1]{\ensuremath{\langle#1\rangle} }

\begin{document}

\preprint{APS/123-QED}

\title{Multiple odd-parity superconducting phases in bilayer transition metal dichalcogenides}

\author{Shota Kanasugi}
\email{kanasugi.shouta.62w@st.kyoto-u.ac.jp}
 \affiliation{%
 Department of Physics, Kyoto University, Kyoto 606-8502, Japan
}%
\author{Youichi Yanase}%
\affiliation{%
 Department of Physics, Kyoto University, Kyoto 606-8502, Japan
}%
\affiliation{%
 Institute for Molecular Science, Okazaki 444-8585, Japan
}%

\date{\today}

\begin{abstract}
We study unconventional superconductivity in a two-dimensional locally noncentrosymmetric triangular lattice. 
The model is relevant to bilayer transition metal dichalcogenides with 2H$_b$ stacking structure, for example. 
The superconducting instability is analyzed by solving the linearized Eliashberg equation within the random phase approximation. 
We show that ferromagnetic fluctuations are dominant owing to the existence of disconnected Fermi pockets near van Hove singularity, and hence odd-parity spin-triplet superconductivity is favored.  
In the absence of the spin-orbit coupling, we find that odd-parity $f$-wave superconducting state is stabilized in a wide range of carrier density and interlayer coupling. 
Furthermore, we investigate impacts of the layer-dependent staggered Rashba and Zeeman spin-orbit coupling on the superconductivity. 
Multiple odd-parity superconducting phase diagrams are obtained as a function of the spin-orbit coupling and Coulomb interaction. 
Especially, a topological chiral $p$-wave pairing state is stabilized in the presence of a moderate Zeeman spin-orbit coupling. 
Our results shed light on a possibility of odd-parity superconductivity in various ferromagnetic van der Waals materials. 
\end{abstract}

\pacs{Valid PACS appear here}
\maketitle


\section{\label{sec1} Introduction}
Searching for odd-parity superconductors, which provide a platform for the intrinsic topological superconductivity \cite{Sato2009,Sato2010,Fu2010}, has been one of central issues in research field of the unconventional superconductivity. 
At present, several solid-state materials are proposed as possible candidates for the odd-parity spin-triplet superconductor, e.g., Sr$_2$RuO$_4$ \cite{Rice1995,Mackenzie2003}, UPt$_3$ \cite{Sauls1994,Tou1998,Joynt2002}, UGe$_2$ \cite{Saxena2000}, URhGe \cite{Aoki2001}, UCoGe \cite{Huy2007}, and UTe$_2$ \cite{Ran2019,Aoki2019}. 
Note that there are now some results conflicting with the spin-triplet pairing in Sr$_2$RuO$_4$ \cite{Yonezawa2014,Kittaka2014,Pustogow2019,Ishida2020}. 
Exploration of spin-triplet superconductivity in systems other than heavy fermions is an important issue. 

There are two important factors for realizing spin-triplet pairing states in solid-state materials, i.e., the ferromagnetic (FM) spin fluctuation and the Fermi surface (FS) structure. 
In the absence of notable FS nesting, the FM fluctuation is enhanced when the Fermi energy lies near the van Hove singularity (vHS). 
Specifically, the so-called type-I\hspace{-.1em}I vHS \cite{Yao2015,Meng2015,Wu2019}, whose saddle points are not located at the time-reversal invariant (TRI) momenta, is preferable for the odd-parity superconductivity. 
On the other hand, a disconnected form of the FS is favorable for the odd-parity pairing since generation of gap nodes is avoidable \cite{Kuroki2001organic,Kuroki2001triangular}. 
Stabilization of odd-parity spin-triplet pairing states has been theoretically proposed in a variety of systems with disconnected FSs, e.g., (TMTSF)$_{2}$X (${\rm X=P}$F$_6$,ClO$_4$) \cite{Kuroki2001organic,Tanaka2004organic,Kuroki2005organic,Nickel2005,Fuseya2005organic}, Na$_x$CoO$_2\cdot y$H$_2$O \cite{Kuroki2004Co,Kuroki2005Co,Ikeda2004Co,Tanaka2004Co,Nisikawa2004Co,Yanase2005multi,Yanase2005role,Mazin2005critical,Mochizuki2005Co}, SrPtAs \cite{Goryo2012,Wang2014SrPtAs}, and doped Kane-Mele model \cite{Fukaya2016}. 

Another intriguing topic for the unconventional superconductivity is relation between crystalline symmetry and the pairing states \cite{Sigrist-Ueda}. 
Particularly, various exotic superconducting (SC) phenomena have been elucidated in locally noncentrosymmetric (NCS) systems \cite{sigrist2014LNCS,maruyama2012locally,Fischer2011,Nakosai2012,Yoshida2012,yoshida2013complex,Yoshida2015,YoshidaDaido2017,Nakamura2017,Ishizuka2018}, in which the inversion symmetry in a local atomic site is broken although the global inversion symmetry is preserved. 
Microscopically, a key aspect of locally NCS systems is the sublattice-dependent antisymmetric spin-orbit coupling (SOC), which leads to exotic superconductivity e.g., singlet-triplet mixed paring states \cite{Fischer2011}, pair density wave states \cite{Yoshida2012,Nakamura2017}, complex stripe states \cite{yoshida2013complex}, and topological superconductivity \cite{Nakosai2012,Yoshida2015,YoshidaDaido2017}. 
Especially, it has been clarified that odd-parity topological superconductivity is realized by a combination of antiferromagnetic spin fluctuations and the sublattice-dependent antisymmetric SOC, namely odd-parity magnetic multipole fluctuations \cite{Ishizuka2018}. 
Thus, it is interesting to study interplay of FM-fluctuation-driven superconductivity and locally NCS crystal structure, in the sense of comparison with the case of the antiferromagnetic-fluctuation-driven superconductivity. 

Considering the above-mentioned aspects, we provide a thorough microscopic investigation of unconventional superconductivity in a two-dimensional (2D) locally NCS triangular lattice [Fig. \ref{fig:lattice}] with disconnected FSs. 
The model is relevant to bilayer transition metal dichalcogenides (TMDs) with 2H$_b$ stacking structure, which is favored in group-VI TMDs  such as MX$_2$ (${\rm M=Mo}$, W and $\mathrm{X}=\mathrm{S}$, Se) \cite{wilson1969,liu2015TMD}. 
In a few layer group-VI TMDs, disconnected FSs are formed around K and K$'$ points owing to the triangular lattice structure. 
Assuming a strong electron correlation, we clarify dominant FM-like spin fluctuations assisted by a type-I\hspace{-.1em}I vHS. 
In fact, ferromagnetism has been recently observed in a few-layer VSe$_2$ \cite{bonilla2018strong} and MnSe$_2$ \cite{o2018room}. 
Since the conduction electrons in TMDs have $d$-orbital character, correlation effects are expected to have considerable impacts on the superconductivity  \cite{Roldan2013,Yuan2014,yuan2015triplet,Hsu2017}. 
We show that odd-parity SC state with $f$-wave symmetry is stabilized by the FM fluctuation in the absence of the SOC. 
On the other hand, the local inversion symmetry breaking in the crystal structure induces layer-dependent staggered Rashba and Zeeman SOC. 
The SOC breaks the spin SU(2) symmetry and lifts the degeneracy of spin-triplet SC states. 
Thus, the SOC controls the internal degree of freedom of odd-parity superconductivity and its topological property. 
We elucidate that multiple odd-parity SC phases with either $p$-wave or $f$-wave pairing, which belong to different irreducible representations (IRs) of the crystal point group, appear by increasing magnitude of the staggered SOC. 
It is shown that the multiple SC phase diagram is a consequence of the selection rule for locally NCS superconductors \cite{Fischer2011} and SOC-induced magnetic anisotropy. 
In addition, topological properties of the stable odd-parity SC states are studied. 
A chiral $p$-wave pairing state in a moderate Zeeman SOC region is identified as a topological SC state in class D.

The rest of the paper is constructed as follows. 
In Sec. \ref{sec2}, we introduce a 2D bilayer triangular lattice Hubbard model including the layer-dependent staggered Rashba and Zeeman SOC. 
The formulation for the microscopic calculations based on the random phase approximation (RPA) and linearized Eliashberg equation is provided. 
In Sec. \ref{sec:fluc}, we study the magnetic fluctuations. 
The dominance of the FM fluctuation and the magnetic anisotropy under the SOC are discussed. 
Numerical results of the Eliashberg equation are shown in Sec. \ref{sec:SC}. 
Stability of $f$-wave SC states is clarified in the absence of the SOC. 
In the presence of the SOC, we identify four stable odd-parity SC states with different pairing symmetry. 
Topological properties of these SC states are also investigated. 
Finally, a brief summary and conclusion are provided in Sec. \ref{sec:summary}.  

\section{\label{sec2} Model and formulation}

\begin{figure}[tbp]
\centering
   \includegraphics[width=85mm,clip]{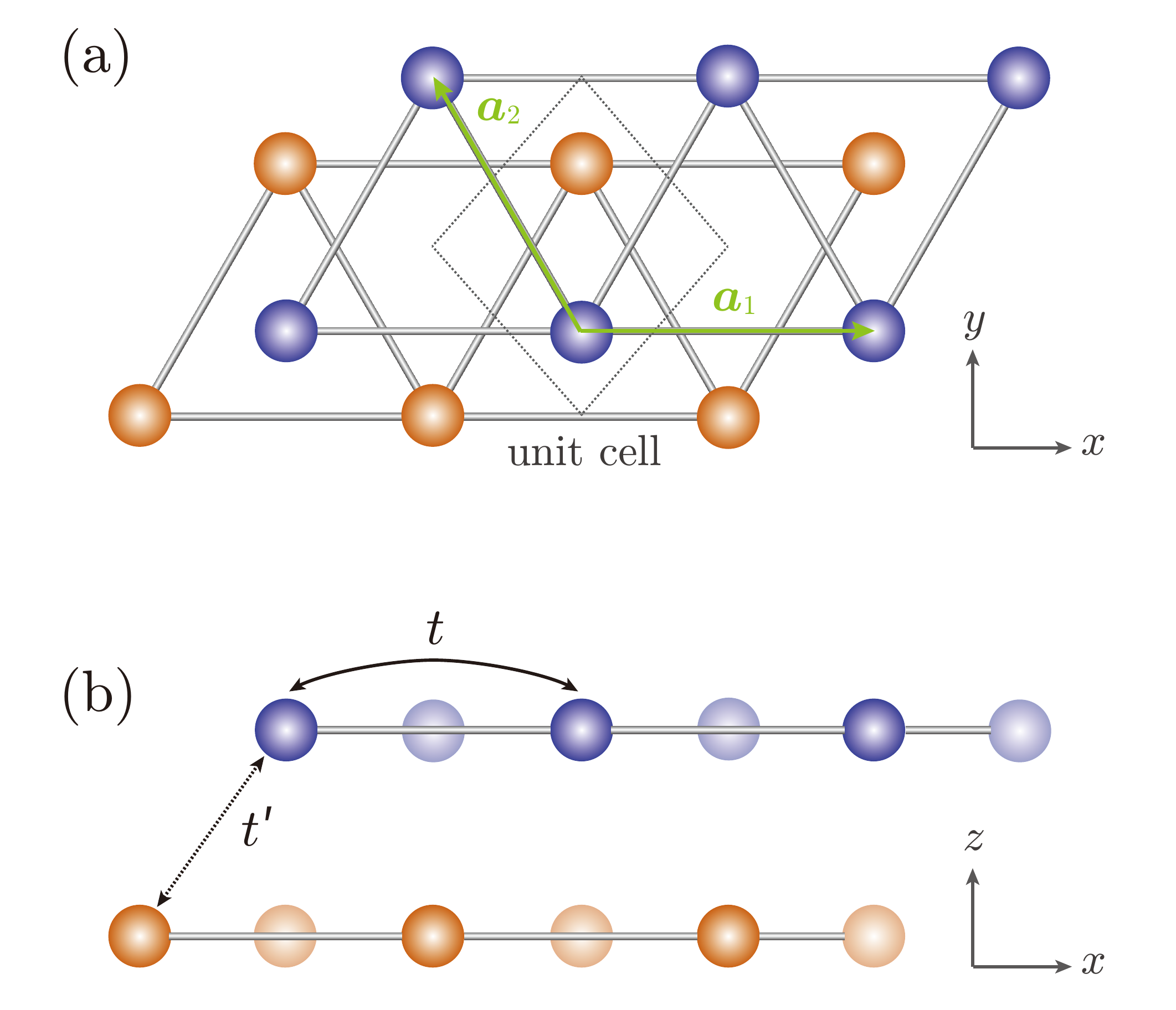}
   \caption{ Crystal structure of the bilayer triangular lattice. (a) is the top view and (b) is the side view. $\bm{a}_{1}$ and $\bm{a}_{2}$ are the lattice vectors. 
$t$ and $t'$ are the intralayer and interlayer hopping integrals, respectively. 
 \label{fig:lattice} }
\end{figure}

We consider a 2D bilayer triangular lattice with the lattice vectors $\bm{a}_1=(1,0)$ and $\bm{a}_2=(-1/2, \sqrt{3}/2)$ [Fig. \ref{fig:lattice}], which is classified into $D_{3d}$ point group. 
The crystal structure is equivalent to that of bilayer TMDs with 2H$_b$ stacking. 
Recently, superconductivity in bilayer MoS$_2$ was realized by symmetric gating \cite{zheliuk2019josephson}. 
On this lattice, we introduce a single-orbital Hubbard model 
$\mathcal{H} = \mathcal{H}_{0} + \mathcal{H}_{\rm int}$. 
The single-particle Hamiltonian $\mathcal{H}_{0}$ with SOC is written as 
\begin{align}
\mathcal{H}_{0} =& \sum_{\bm{k},m,s} \left(\varepsilon(\bm{k})-\mu\right) c_{\bm{k},ms}^{\dag} c_{\bm{k},ms} \nonumber\\
& + \sum_{\bm{k},s} \left( \eta(\bm{k}) c_{\bm{k},as}^{\dag} c_{\bm{k},bs} + \mathrm{H.c.} \right) \nonumber\\
&+ \sum_{\bm{k},\zeta,\zeta'} \sum_{j=1,2}
\alpha_{j} \bm{g}_{j}(\bm{k}) \cdot \bm{\sigma}_{ss'} \tau_{mm'}^{z} c_{\bm{k},ms}^{\dag} c_{\bm{k},ms'}, 
\end{align}
where $c_{\bm{k},ms}$ is the annihilation operator for an electron with momentum $\bm{k}$ and spin $s=\uparrow,\downarrow$ on layer $m=a,b$. 
$\zeta=(m,s)$ is the abbreviated notation, and $\sigma^{\mu}$ ($\tau^{\nu}$) is the Pauli matrix for spin (layer) degrees of freedom. 
The first term is the kinetic energy term. 
The single-electron kinetic energy is described as 
\begin{align}
\varepsilon(\bm{k}) &= 2t \left[ \cos\bm{k}\cdot\bm{a}_1 + \cos\bm{k}\cdot\bm{a}_2 + \cos\bm{k}\cdot(\bm{a}_1+\bm{a}_2) \right], 
\end{align}
by taking into account the nearest-neighbor hopping. 
We choose the hopping integral $t$ as a unit of energy ($t=1$). 
The chemical potential $\mu$ is determined to fix the carrier density as $n$. 
The second term is the interlayer coupling. 
The interlayer hybridization function is given by
\begin{align}
\eta(\bm{k}) &= t' \left[1+e^{-i\bm{k}\cdot\bm{a}_2}+e^{-i\bm{k}\cdot(\bm{a}_1+\bm{a}_2)} \right]. 
\end{align}
In this study, we assume that the interlayer hopping integral $t'$ is smaller than the intralayer hopping integral $t$ (i.e., $t'<t$). 
The third term is the layer-dependent staggered SOC, which is originated from the locally NCS crystal structure and a spin-dependent intralayer hopping. 
Since the local site symmetry is $C_{3v}$, the $g$-vector $\bm{g}_{j}(\bm{k})$ should belong to $A_{2u}$ IR of $D_{3d}$ which becomes trivial $A_1$ IR in $C_{3v}$ [see Table \ref{tab:basis}]. 
In this study, we consider two kinds of $g$-vectors as
\begin{align}
\bm{g}_{1}(\bm{k}) &= \frac{1}{\Lambda}\left[ 
\frac{\sqrt{3}}{2}\left\{ \sin\bm{k}\cdot(\bm{a}_1+\bm{a}_2)+\sin\bm{k}\cdot\bm{a}_{2}\right\} \hat{\bm{x}} \right. \nonumber\\
&\left.  
-\left\{ \sin\bm{k}\cdot\bm{a}_{1}
+\frac{\sin\bm{k}\cdot(\bm{a}_{1}+\bm{a}_2)-\sin\bm{k}\cdot\bm{a}_{2}}{2} \right\} \hat{\bm{y}}
\right] , \label{eq:Rashba} \\
\bm{g}_{2}(\bm{k}) &= \frac{2}{3\sqrt{3}}\left[
\sin\bm{k}\cdot\bm{a}_1 + \sin\bm{k}\cdot\bm{a}_2 - \sin\bm{k}\cdot(\bm{a}_1+\bm{a}_2) \right] \hat{\bm{z}} ,
\label{eq:Zeeman}
\end{align}
where $\Lambda=1.7602$. 
Equations (\ref{eq:Rashba}) and (\ref{eq:Zeeman}) are the Rashba and Zeeman SOC, respectively. 
Both terms belong to $A_{2u}$ IR. 
The Rashba (Zeeman) SOC originates from the out-of-plane (in-plane) local inversion symmetry breaking at each layers. 
The constant factors are chosen as $\mathrm{Max}_{\bm{k}}|\bm{g}_{j}(\bm{k})|=1$. 
Although the Rashba SOC is negligible compared to the Zeeman SOC in some TMDs \cite{saito2016superconductivity,Nakamura2017}, we treat both of them on equal footing to provide a general calculation not limited to existing TMDs. 
The on-site Coulomb interaction is given by
\begin{align}
\mathcal{H}_{\rm int} &= U \sum_{\bm{i},m} n_{\bm{i},m\uparrow} n_{\bm{i},m\downarrow},
\end{align}
where $n_{\bm{i},ms}=c_{\bm{i},ms}^{\dag}c_{\bm{i},ms}$ is the electron density operator on site $\bm{i}$. 
Strong repulsive electron-electron interaction may be present owing to the $d$-orbital character of conduction carries in TMDs. 
We treat $\mathcal{H}_{\rm int}$ in the RPA. 

We study the SC instability in this model by solving the linearized Eliashberg equation  
\begin{eqnarray}
\lambda \Delta_{\zeta\zeta'}(k) =&& -\frac{T}{N}\sum_{k'}\sum_{\{\zeta_j\}} V_{\zeta\zeta_1,\zeta_2\zeta'}(k-k') \nonumber\\
&&\times G_{\zeta_3\zeta_1}(-k')\Delta_{\zeta_3\zeta_4}(k')G_{\zeta_4\zeta_2}(k'), 
\end{eqnarray}
where we used the abbreviated notation $k=(\bm{k}, i\omega_{p})$ and $\omega_p=(2p+1)\pi T$ is the fermionic Matsubara frequency. 
The noninteracting temperature Green's function is given by 
$\hat{G}(k)=[i\omega_p\hat{1}-\hat{\mathcal{H}}_{0}(\bm{k})]^{-1}$. 
$\lambda$ and $\hat{\Delta}(k)$ are the eigenvalue and gap function, respectively. 
In the RPA, the effective pairing interaction $\hat{V}(q)$ can be described as 
\begin{align}
\hat{V}(q)=-\hat{\Gamma}^{(0)}\hat{\chi}(q)\hat{\Gamma}^{(0)}-\hat{\Gamma}^{(0)}, 
\end{align}
by using the RPA susceptibility
\begin{align}
\hat{\chi}(q)=\hat{\chi}^{(0)}(q)[\hat{1}-\hat{\Gamma}^{(0)}\hat{\chi}^{(0)}(q)]^{-1}. 
\end{align}
Here, the irreducible susceptibility is defined as 
\begin{align}
\chi^{(0)}_{\zeta_1\zeta_2,\zeta_3\zeta_4}(q) = -\frac{T}{N}\sum_{k}G_{\zeta_3\zeta_1}(k)G_{\zeta_2\zeta_4}(k+q).
\end{align}
The bare irreducible vertex in this model is obtained as 
\begin{align}
\Gamma^{(0)}_{\zeta_1\zeta_2, \zeta_3\zeta_4}=&\frac{U}{2}\delta_{m_1m_2}\delta_{m_3m_4}\delta_{m_1m_3} \nonumber\\
&\times(\bm{\sigma}_{s_1s_2}\cdot\bm{\sigma}_{s_4s_3}-\delta_{s_1s_2}\delta_{s_4s_3}). 
\end{align}
In the following numerical calculations, we set $T/t=0.02$,  $64\times64$ $\bm{k}$-points, and 1024 Matsubara frequencies.

\section{Magnetic fluctuation \label{sec:fluc}}
\begin{figure}[tbp]
\centering
   \includegraphics[width=90mm,clip]{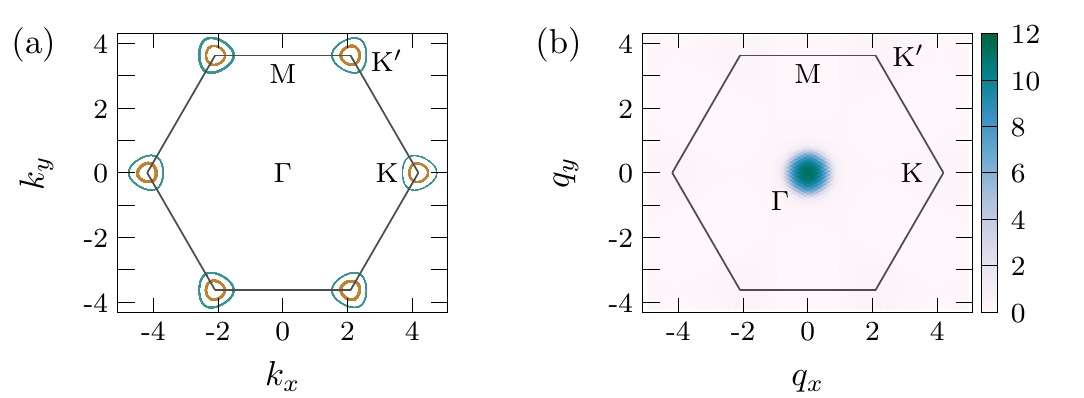}
   \caption{(a) Fermi surfaces for $t'/t=0.2$, $n=0.1$ and $\alpha_{j}=0$. (b) Momentum dependence of the intralayer magnetic susceptibility $\chi_{aa(=bb)}^{S}(\bm{q},0)$ for $t'/t=0.2$, $n=0.1$, $\alpha_{j}=0$, and $U=5.0$. 
 \label{fig:chi_alpha0} }
\end{figure}

In this section, we study the magnetic fluctuation by introducing magnetic susceptibilities as follows:
\begin{align}
\chi_{mm'}^{\mu\nu}(q)=\sum_{\{s_j\}}\sigma_{s_1s_2}^{\mu}\chi_{ms_1 ms_2, m's_3 m's_4}(q)\sigma_{s_4s_3}^{\nu} , 
\end{align}
where $\mu, \nu = x, y, z$. 
The magnetic fluctuation parallel (perpendicular) to the $c$-axis is characterized by $\chi^{\parallel}\equiv\chi^{zz}$ ($\chi^{\perp}\equiv(\chi^{xx}+\chi^{yy})/2$). 
In the following, we consider low doping regimes, in which small disconnected Fermi pockets are formed around the K and  K' points [see Fig. \ref{fig:chi_alpha0}(a)]. 
This condition is relevant to electron-doped bilayer TMDs with 2H$_b$ stacking structure \cite{zheliuk2019josephson}. 

\begin{figure}[tbp]
\centering
   \includegraphics[width=86mm,clip]{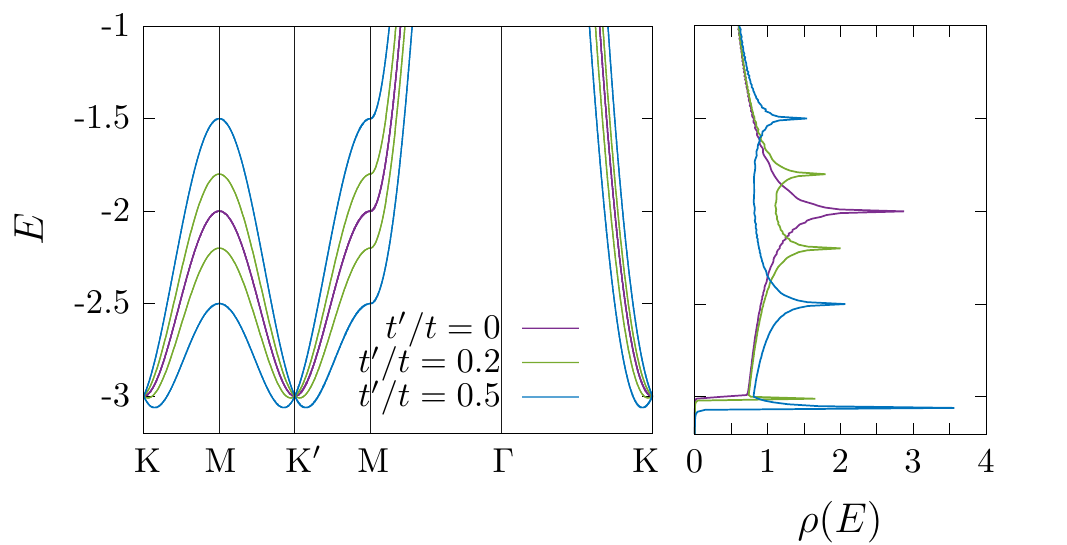}
   \caption{Band structure and density of states for $\alpha_{j}=0$ and $\mu=0$. 
 \label{fig:band_dos} }
\end{figure}

First, we investigate the magnetic fluctuations in the absence of the SOC. 
In this case, there is no magnetic anisotropy, and hence $\chi^{\parallel}=\chi^{\perp}(\equiv\chi^{S})$.  
In Fig. \ref{fig:chi_alpha0}(b), we show momentum dependence of the intralayer magnetic susceptibility $\chi_{aa}^{S}(=\chi_{bb}^{S})$ for $t'/t=0.2$, $n=0.1$ and $\alpha_j=0$. 
The magnetic susceptibility is sharply peaked at $\bm{q}\simeq\bm{0}$, indicating dominant FM fluctuation in this system. 
The FM fluctuation is partially owing to the smallness of the FS.  
Besides, the FM fluctuation is enhanced because the Fermi level lies near the type-I\hspace{-.1em}I vHS, which is located slightly away from the K (K$'$) point. 
This type-I\hspace{-.1em}I vHS originates from the band splitting at the band bottom due to a finite interlayer coupling, and hence it is a fingerprint of the bilayer structure. 
In the 2H$_b$ stacking, the interlayer hybridization vanishes at the K (K$'$) points as ensured by the threefold rotational symmetry \cite{liu2015TMD,Akashi2015,Akashi2017}. 
Therefore, Dirac-type linear dispersion appears around the K (K$'$) point [see Fig. \ref{fig:band_dos}], and it gives rise to the type-I\hspace{-.1em}I vHS similar to the Rashba model \cite{Kanasugi2018}. 
Indeed, Fig. \ref{fig:band_dos} reveals a large density of states near the band bottom. 

\begin{figure}[tbp]
\centering
   \includegraphics[width=90mm,clip]{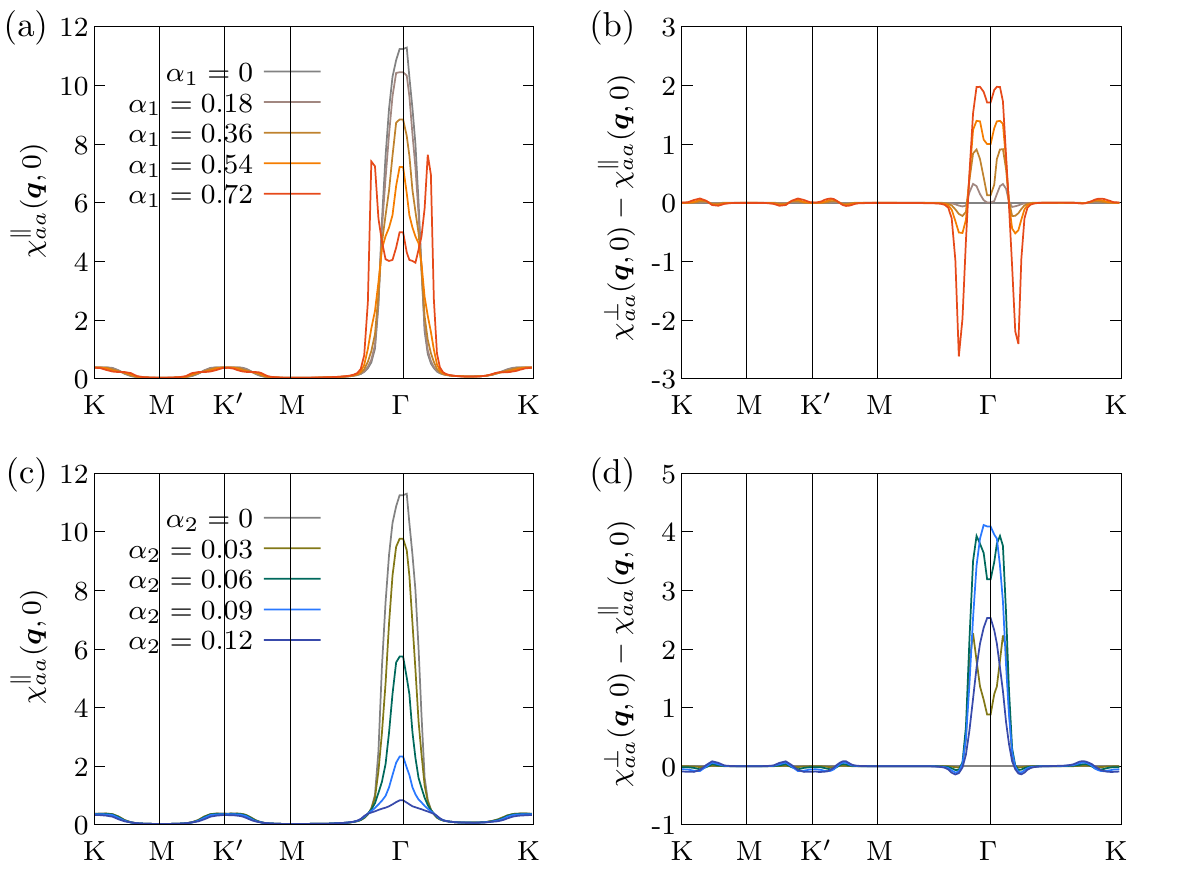}
   \caption{Momentum dependence of the magnetic susceptibilities along the symmetry axis for several values of the (a), (b) Rashba SOC $\alpha_{1}$ and (c), (d) Zeeman SOC $\alpha_{2}$. 
Parameters are set to be $t'/t=0.2$, $n=0.1$, $U=5.0$, and $T=0.02$. 
(a), (c) An intralayer component $\chi_{aa}^{\parallel}(\bm{q},0)$, and (b), (d) the anisotropy $\chi_{aa}^{\perp}(\bm{q},0)-\chi_{aa}^{\parallel}(\bm{q},0)$. 
 \label{fig:chi_SOC} }
\end{figure}

Next, we show the magnetic fluctuations in the presence of the SOC. 
In locally NCS systems, a sublattice-dependent staggered SOC gives a significant impact on the electronic structure when the ratio of the SOC and the intersublattice coupling is large \cite{maruyama2012locally}. 
Since the interlayer coupling $\eta(\bm{k})$ vanishes at the K (K$'$) point \cite{liu2015TMD,Akashi2015,Akashi2017}, the ratio $\varphi_{j}(\bm{k})\equiv|\alpha_{j}\bm{g}_{j}(\bm{k})|/|t'\eta(\bm{k})|$ can be large on the FS. 
Hence, the magnetic fluctuation is strongly affected by the staggered SOC. 
The SOC dependences of the magnetic susceptibilities are shown in Fig.  \ref{fig:chi_SOC}. 
The sharp peak of the magnetic susceptibility at the $\Gamma$ point is gradually suppressed by increasing $\alpha_j$ [Figs. \ref{fig:chi_SOC}(a) and \ref{fig:chi_SOC}(c)], and the FM fluctuation is weakened. 
The suppression of the FM fluctuation is significant in the case of the Zeeman SOC, since the ratio of the SOC and interlayer coupling has a larger value than that in the case of the Rashba SOC [i.e., $\varphi_{2}(\bm{k}_{\rm F})>\varphi_{1}(\bm{k}_{\rm F})$]. 
Figures \ref{fig:chi_SOC}(b) and \ref{fig:chi_SOC}(d) reveal appearance of the magnetic anisotropy ($\chi^{\parallel}\neq\chi^{\perp}$) owing to the violation of the spin rotational symmetry. 
The Rashba SOC monotonically increases the magnetic anisotropy mainly at around the $\Gamma$ point [Fig. \ref{fig:chi_SOC}(b)]. 
On the other hand, the growth of the magnetic anisotropy by the Zeeman SOC is nonmonotonic [Fig. \ref{fig:chi_SOC}(d)]. 
Although the SOC dependence of the magnetic anisotropy is complicated, we note that $\chi^{\perp}>\chi^{\parallel}$ is always realized at the $\Gamma$ point. 
Thus, a FM-like magnetic structure with an in-plane spin-alignment is favored in the presence of the SOC. 
It should be noticed that such an in-plane FM ordering has been observed in atomically thin film of group-V TMD VSe$_2$ \cite{bonilla2018strong}. 
The superconductivity is significantly affected by this magnetic anisotropy as we demonstrate in Sec. \ref{sec:SC_SOC}. 

\section{Superconductivity \label{sec:SC}}
Here, we illustrate numerical results of the Eliashberg equation in the framework of the RPA. 
Multiple odd-parity SC phases stabilized by FM fluctuations are demonstrated.

\begin{figure}[tbp]
\centering
   \includegraphics[width=80mm,clip]{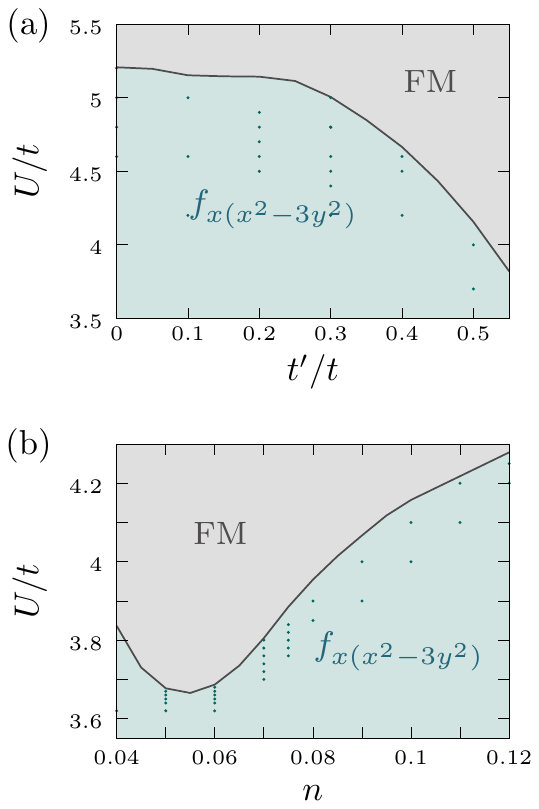}
   \caption{(a) Phase diagram for the Coulomb interaction $U$ and interlayer hopping $t'$ at $n=0.1$ and $T=0.02$. (b) Phase diagram for the Coulomb interaction $U$ and carrier density $n$ at $t'/t=0.5$ and $T=0.02$. The solid line is the phase boundary between FM-like ordered state and paramagnetic state, at which the Stoner factor $S=\mathrm{max}[\hat{\Gamma}^{(0)}\hat{\chi}(q)]$ becomes unity. In the paramagnetic phase, the $f_{x(x^2-3y^2)}$-wave pairing state is stabilized. 
 \label{fig:phase_alpha0} }
\end{figure}

\subsection{Superconductivity without spin-orbit coupling \label{sec:SC_alpha0}}
First, we show the SC phases in the absence of the SOC. 
Figure \ref{fig:phase_alpha0}(a) (Figure \ref{fig:phase_alpha0}(b)) shows phase diagrams as a function of the interlayer hopping $t'$ (carrier density $n$) and Coulomb interaction $U$ at $n=0.1$ ($t'/t=0.5$). 
Owing to the dominant FM spin fluctuations, odd-parity spin-triplet $f_{x(x^2-3y^2)}$-wave SC states, which are classified into $A_{2u}$ or $E_{u}$ IRs in the presence of the SOC, are stabilized in the whole parameter region.  
This $f_{x(x^2-3y^2)}$-wave SC state is a full gap state and mainly caused by intralayer Cooper pairing. 
The gap function for the $f_{x(x^2-3y^2)}$-wave SC state is illustrated in Fig. \ref{fig:gap_A2u} (a). 
Since the effective pairing interaction for spin-triplet superconductivity is approximated as $V^{\rm triplet}\simeq-(U^2/2)\chi^{S}$ in the absence of SOC, the magnetic fluctuation favors the gap function with the same sign on each pieces of the FS connected by a vector $\bm{Q}$. 
Here, $\bm{Q}$ is the wave vector at which the magnetic susceptibility is enhanced. 
As shown in Fig. \ref{fig:chi_alpha0}(b), the magnetic susceptibility is sharply peaked at $\bm{q}\simeq\bm{0}$ (i.e., $\bm{Q}\simeq\bm{0}$). 
Thus, the $f_{x(x^2-3y^2)}$-wave SC state is stabilized to avoid generation of gap nodes at the $\mathrm{K}$ and $\mathrm{K}'$ points \cite{Kuroki2001organic,Kuroki2001triangular}.

\subsection{Superconductivity and spin-orbit coupling \label{sec:SC_SOC}}
Next, we investigate superconductivity in the presence of the layer-dependent staggered SOC. 
In the following discussion, we describe the SC gap function as 
$\Delta_{ms,m's'}^{i}(k)=\sum_{\mu\nu}d^{\mu\nu}_{i}(k)\bar{\sigma}^{\mu}_{ss'}\tau^{\nu}_{mm'}$, 
where $i=1,2$ is the index for 2D IRs and $\bar{\sigma}_{ss'}^{\mu}=[\sigma^{\mu}i\sigma^y]_{ss'}$ ($\mu=0, x, y, z$). 
In the presence of the SOC, symmetry of SC states is classified based on the crystallographic point group. 
Then, the gap function belongs to one of IRs of $D_{3d}$ point group shown in Table \ref{tab:basis}. 
The SC instability is discussed by solving the Eliashberg equation under symmetry constraint for each of the IRs (see Appendix \ref{append:symmetry}). 

\begin{table}[tbp] 
\caption{\label{tab:basis}
2D basis gap functions for the IRs of trigonal $D_{3d}$ point group. The second column shows the compatibility relations between $D_{3d}$ and $C_{3v}$. 
}
\centering
\begin{ruledtabular}
{\renewcommand \arraystretch{1.3}
 \begin{tabular}{cccc} 
	 $D_{3d}$ & $D_{3d} \downarrow C_{3v}$ & Basis functions with $k_z=0$ \\ \hline
  $A_{1g}$ & $A_1$ & $1$ \\ 
  $A_{2g}$ & $A_2$ & $k_xk_y(k_x^2-3k_y^2)(3k_x^2-k_y^2)$ \\ 
  $E_{g}$ & $E$ & $\{ k_x k_y, k_x^2-k_y^2\}$ \\ 
  $A_{1u}$ & $A_2$ & $k_x\hat{\bm{x}}+k_y\hat{\bm{y}}$, $k_y(3k_x^2-k_y^2)\hat{\bm{z}}$ \\ 
  $A_{2u}$ & $A_1$ & $k_x\hat{\bm{y}}-k_y\hat{\bm{x}}$, $k_x(k_x^2-3k_y^2)\hat{\bm{z}}$ \\ 
  $E_{u}$ & $E$ & $\{k_x\hat{\bm{y}}+k_y\hat{\bm{x}}, k_x\hat{\bm{x}}-k_y\hat{\bm{y}}\}$, $\{ k_x\hat{\bm{z}}, k_y\hat{\bm{z}} \}$ \\
\end{tabular}
}
\end{ruledtabular}
\end{table} 

\begin{figure}[tbp]
\centering
   \includegraphics[width=90mm,clip]{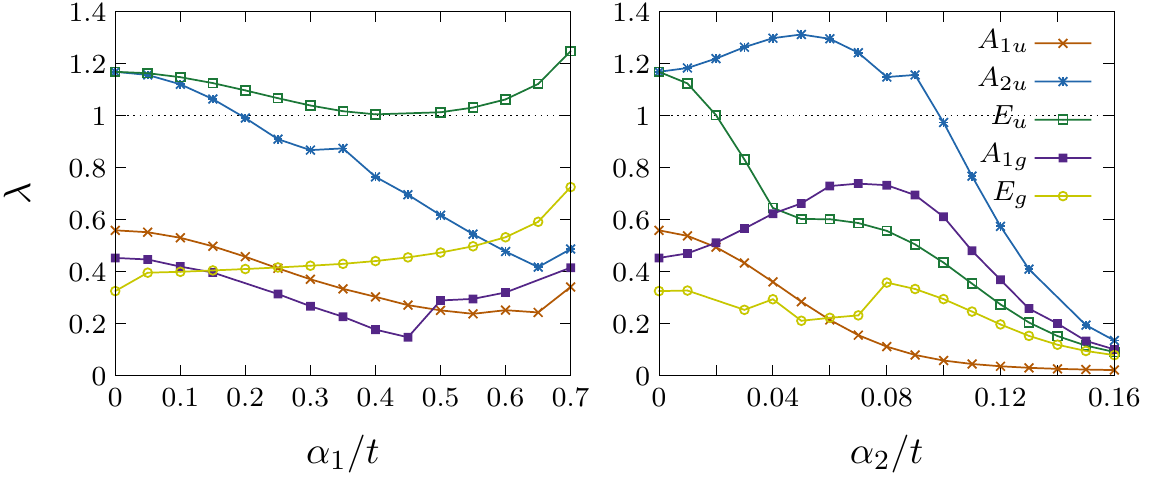}
   \caption{SOC dependence of eigenvalues of the Eliashberg equation $\lambda$ at $t'/t=0.2$, $n=0.1$, $U=4.8$, and $T=0.02$. 
The eigenvalues for $A_{1u}$ ($p$-wave), $A_{2u}$ ($f_{x^2(x^2-3y^2)}$-wave), $E_{u}$ ($f_{x^2(x^2-3y^2)}$-wave or $p$-wave), $A_{1g}$ ($s$-wave), and $E_g$ ($d$-wave) pairing states are illustrated. 
 \label{fig:lambda_SOC} }
\end{figure}

\begin{figure*}[htbp]
\centering
   \includegraphics[width=175mm,clip]{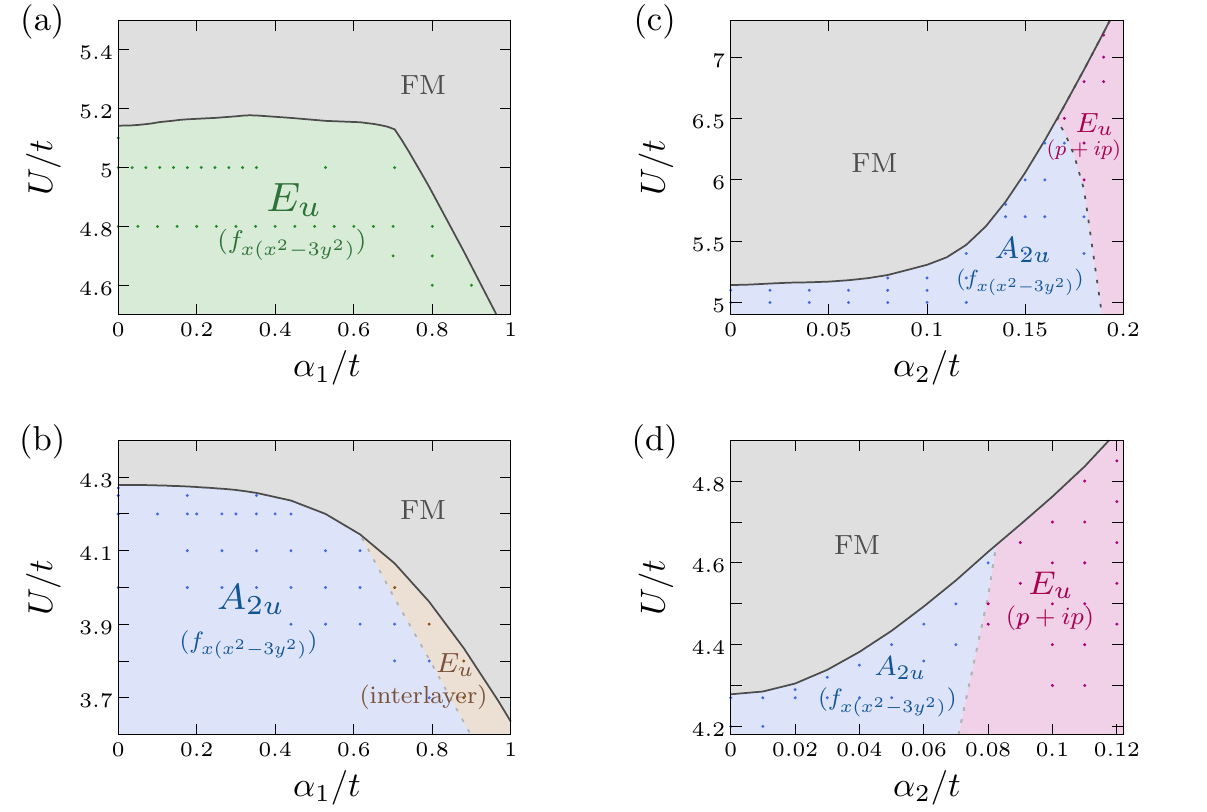}
   \caption{(a) and (b) Phase diagram for the Coulomb interaction $U$ and staggered Rashba SOC $\alpha_1$ at $\alpha_2=0$ and $T=0.02$. 
(c) and (d) Phase diagram for the Coulomb interaction $U$ and staggered Zeeman SOC $\alpha_2$ at $\alpha_1=0$ and $T=0.02$. 
(a) and (c) $t'/t=0.2$ and $n=0.1$, and (b) and (d) $t'/t=0.5$ and $n=0.12$. 
The solid line is the phase boundary between FM-like ordered state and paramagnetic state. 
In the paramagnetic phase, the odd-parity $A_{2u}$ and $E_{u}$ SC states are illustrated. 
 \label{fig:phase_SOC} }
\end{figure*}

Figure \ref{fig:lambda_SOC} shows the SOC dependence of eigenvalues of the Eliashberg equation $\lambda$ at $t'/t=0.2$, $n=0.1$, and $U=4.8$. 
Owing to the dominant FM fluctuation, the intralayer $f_{x(x^2-3y^2)}$-wave ($A_{2u}$ or $E_{u}$) pairing state is predominant and the $p$-wave ($A_{1u}$, $A_{2u}$, or $E_u$) pairing state is subdominant for $\alpha_j=0$. 
On the other hand, the eigenvalues of even-parity $s$-wave ($A_{1g}$) and $d$-wave ($E_g$) pairing states are smaller than those of odd-parity pairing states. 
The eigenvalues of the $A_{2u}$ and $E_{u}$ SC states are equal at $\alpha_j=0$, since the spin part of the gap function is threefold degenerated in the absence of the SOC. 
By turning on the staggered Rashba (Zeeman) SOC, the degeneracy is lifted due to violation of the spin-rotational symmetry, and the $E_{u}$ ($A_{2u}$) SC state is stabilized as $\lambda^{A_{2u}}<\lambda^{E_u}$ ($\lambda^{A_{2u}}>\lambda^{E_u}$). 
For these parameters, the spin direction of the SC state is determined by the selection rule for locally NCS superconductors \cite{Fischer2011,Ishizuka2018}, which originates from a modulation of the one-particle Green's function by the staggered SOC. 
According to the selection rule, spin-singlet state and spin-triplet state with $\bm{d}(\bm{k})\parallel\bm{g}(\bm{k})$ are stable for intralayer pairing, while only spin-triplet state with $\bm{d}(\bm{k})\perp\bm{g}(\bm{k})$ is stable for interlayer pairing. 
In a small SOC region, the leading order parameter for the $A_{2u}$ ($E_{u}$) pairing state possesses the intralayer $f_{x(x^2-3y^2)}$-wave symmetry with $\bm{d}\parallel\hat{\bm{z}}$ ($\bm{d}\parallel\hat{\bm{x}}, \hat{\bm{y}}$). 
Thus, the $A_{2u}$ ($E_u$) SC state is destabilized by the staggered Rashba (Zeeman) SOC, since the leading intralayer order parameter with $\bm{d}\perp\bm{g}_{1}$ ($\bm{d}\perp\bm{g}_{2}$) is incompatible with the selection rule. 
In addition, to be compatible with the selection rule, the gap function is modified in a large SOC region. 
For example, the leading order parameter of the $E_u$ pairing state exhibits $p$-wave symmetry for $\alpha_2/t\gtrsim0.04$, while that shows $f_{x(x^2-3y^2)}$-wave symmetry for $\alpha_2/t\lesssim0.04$ [see right panel of Fig. \ref{fig:lambda_SOC}]. 
As demonstrated above, competition of various SC states with different pairing symmetry can be controlled by the staggered SOC. 

\begin{table*}[tbp] 
\caption{\label{tab:gap}
Leading order parameters for the odd-parity $A_{2u}$ and $E_{u}$ SC states. 
$\Delta^s(k)$, $\Delta^{p_x}(k)$, $\Delta^{p_y}(k)$, and $\Delta^f(k)$ denote gap functions which possess momentum dependence with $s$-wave, $p_x$-wave, $p_y$-wave, and $f_{x(x^2-3y^2)}$-wave symmetry. 
The third column shows the phase diagram in which the corresponding SC state is stabilized. 
The last column is figures which illustrate the gap functions. 
}
\centering
\begin{ruledtabular}
{\renewcommand \arraystretch{1.3}
 \begin{tabular}{ccccc} 
	 IR & Leading order parameter & Phase diagram & Gap function  \\ \hline
$A_{2u}$ & $\Delta^f(k)\bar{\sigma}^z\tau^0 + \alpha_{j} \Delta^s(k) \bar{\sigma}^0\tau^z$ & Figs. \ref{fig:phase_SOC}(b), \ref{fig:phase_SOC}(c), and \ref{fig:phase_SOC}(d) & Fig. \ref{fig:gap_A2u} \\
   &  $\{ \Delta^f(k)\bar{\sigma}^y\tau^0, \Delta^f(k)\bar{\sigma}^x\tau^0 \}$ & Fig. \ref{fig:phase_SOC}(a) & Figs. \ref{fig:gap_Eu}(a) and \ref{fig:gap_Eu}(b) \\
$E_{u}$ & $\{ \Delta^{p_{x}}(k)\bar{\sigma}^z\tau^0, \Delta^{p_{y}}(k)\bar{\sigma}^z\tau^0 \}$ & Figs. \ref{fig:phase_SOC}(c) and \ref{fig:phase_SOC}(d) & Figs. \ref{fig:gap_Eu}(c) and \ref{fig:gap_Eu}(d) \\
   &  $\{ \Delta^{f}(k)\bar{\sigma}^z\tau^x, \Delta^{s}(k)\bar{\sigma}^z\tau^y \}$ & Fig. \ref{fig:phase_SOC}(b) & Figs. \ref{fig:gap_Eu}(e) and \ref{fig:gap_Eu}(f) \\
\end{tabular}
}
\end{ruledtabular}
\end{table*} 

\begin{figure}[tbp]
\centering
   \includegraphics[width=85mm,clip]{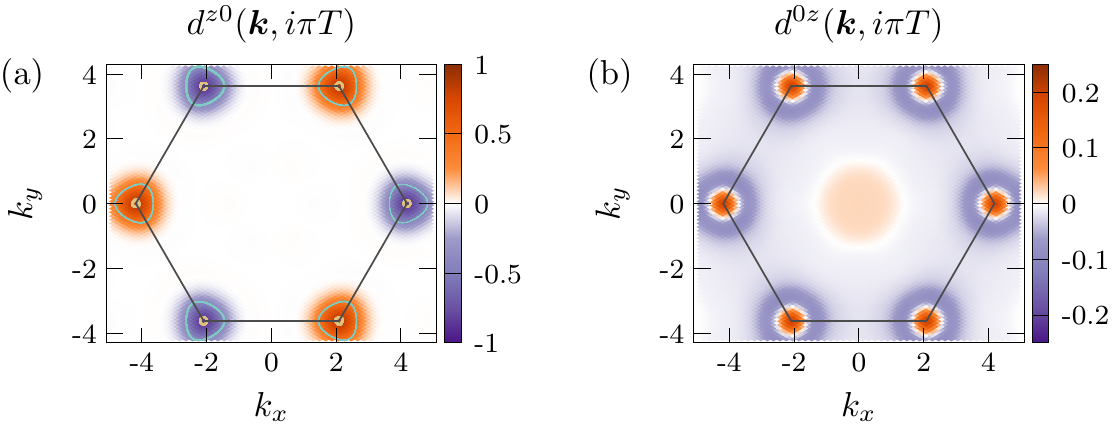}
   \caption{Gap functions for the $A_{2u}$ SC state at $t'/t=0.2$, $n=0.1$, $\alpha_{2}=0.1$, and  $U=5.2$. 
(a) Leading intralayer $f_{x(x^2-3y^2)}$-wave component $d^{z0}(\bm{k}, i\pi T)$ and (b) parity-mixing-induced $s$-wave component $d^{0z}(\bm{k}, i\pi T)$. 
The gap functions are normalized so that the maximum amplitude of the leading order parameter becomes unity. 
Corresponding FS is illustrated in the left panel. 
Eigenvalues of the Eliashberg equation is $\lambda=2.83905$. 
 \label{fig:gap_A2u} }
\end{figure}

\begin{figure}[tbp]
\centering
   \includegraphics[width=85mm,clip]{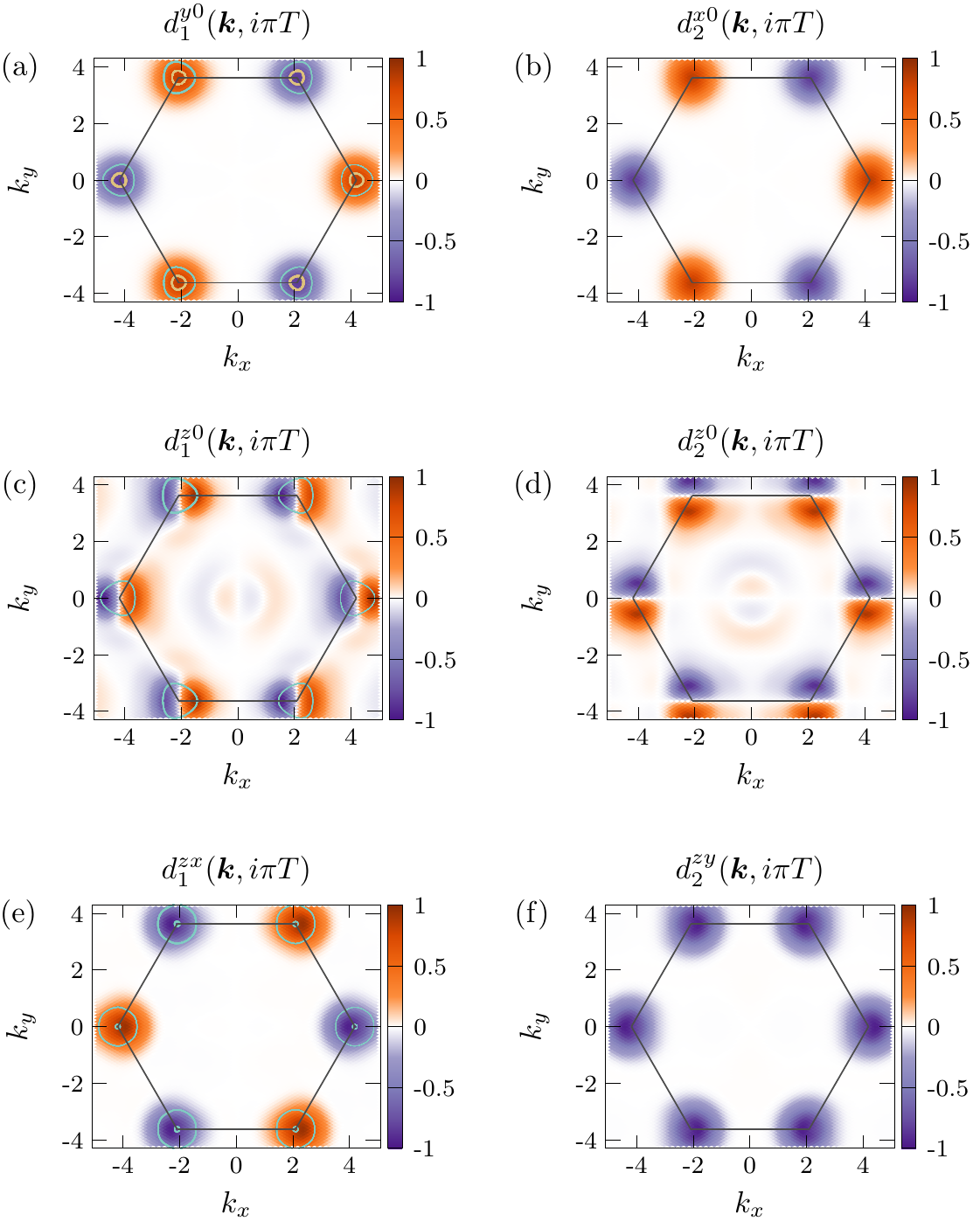}
   \caption{Leading components of the gap function for the $E_{u}$ SC states.  (a), (b) $t'/t=0.2$, $n=0.1$, $\alpha_{1}=0.35204$, and  $U=5.0$. 
(c), (d) $t'/t=0.2$, $n=0.1$, $\alpha_{2}=0.19$, and  $U=7.18$. 
(e), (f) $t'/t=0.5$, $n=0.12$, $\alpha_{1}=0.8801$, and  $U=3.8$. 
The gap functions are normalized so that the maximum amplitude of the leading order parameter becomes unity. 
Corresponding FSs are illustrated in the left panels. 
Eigenvalues of the Eliashberg equation is $\lambda=2.1145, 1.04829
, 2.32491$ in (a,b), (c,d), and (e,f), respectively. 
 \label{fig:gap_Eu} }
\end{figure}

Figure \ref{fig:phase_SOC} shows phase diagrams as a function of the staggered SOC $\alpha_{j}$ and Coulomb interaction $U$. 
We found that an odd-parity SC state with either $A_{2u}$ or $E_{u}$ symmetry is stabilized and it is controlled by magnitude of the SOC and Coulomb interaction. 
The gap functions for these odd-parity SC states are illustrated in Table \ref{tab:gap} and Figs. \ref{fig:gap_A2u} and \ref{fig:gap_Eu}. 
It should be noticed the Zeeman SOC significantly affects the superconductivity compared to the Rashba SOC because the Zeeman SOC takes a large magnitude near the K point. 
Therefore, superconductivity in a trigonal system with in-plane inversion symmetry breaking is affected by a moderate SOC. 

In the presence of the Rashba SOC, the superconductivity exhibits different behaviors depending on the magnitude of the interlayer hopping. 
In the case of a small interlayer hopping $t'/t=0.2$, the staggered Rashba SOC stabilizes only the $E_{u}$ SC state [Fig. \ref{fig:phase_SOC}(a)], whose leading order parameters are intralayer spin-triplet components $\{d^{y0}_{1}, d^{x0}_{2}\}$ with $f_{x(x^2-3y^2)}$-wave symmetry [Figs. \ref{fig:gap_Eu}(a) and \ref{fig:gap_Eu}(b)]. 
This $E_{u}$ $f_{x(x^2-3y^2)}$-wave SC state is compatible with the selection rule as we already demonstrate for Fig. \ref{fig:lambda_SOC}. 
On the other hand, in the case of a large interlayer hopping $t'/t=0.5$, the $A_{2u}$ or $E_u$ SC states are stabilized depending on the magnitude of the Rashba SOC [Fig. \ref{fig:phase_SOC}(b)]. 
The $A_{2u}$ SC state is favored for a small Rashba SOC region ($0\lesssim\alpha_1/t\lesssim0.8$), while the $E_u$ SC state is favored for a large Rashba SOC region ($\alpha_1/t\gtrsim0.8$). 
This multiple SC phase diagram is a consequence of competition between the selection rule and magnetic anisotropy. 
The $A_{2u}$ SC state with the $f_{x(x^2-3y^2)}$-wave leading order parameter $d^{z0}$ is incompatible with the selection rule because $\bm{d}\perp\bm{g}_{1}$ in the whole Brillouin zone. 
The stabilization of the $A_{2u}$ SC state may be attributed to the magnetic anisotropy. 
The magnetic anisotropy under the Rashba SOC is always $\chi^{\perp}>\chi^{\parallel}$ near the $\Gamma$ point like that for $t'/t=0.2$ [see Fig. \ref{fig:chi_SOC}(b)]. 
Since the effective pairing interaction for the spin-triplet pair amplitude $d^{z\nu}$ can be approximated as $V^{\rm eff}\approx-(U^2/2)(2\chi^{\perp}-\chi^{\parallel})$, the magnetic anisotropy $\chi^{\perp}>\chi^{\parallel}$ favors the spin-triplet pairing with $\bm{d}\parallel\hat{\bm{z}}$. 
Thus, the $A_{2u}$ SC state is stabilized contrary to the selection rule. 
Note that impacts of a sublattice-dependent staggered SOC on the electronic structure are generally weakened by increasing the intersublattice coupling \cite{maruyama2012locally}. 
Leading order parameter of the $E_{u}$ SC state for $\alpha_1/t\gtrsim0.8$ is interlayer spin-triplet components $\{d^{zx}_1, d^{zy}_2\}$ [Figs. \ref{fig:gap_Eu}(e) and \ref{fig:gap_Eu}(f)], which are compatible with the selection rule. 
The enhancement of the interlayer order parameters $\{d^{zx}_1, d^{zy}_2\}$ is attributed to the large interlayer coupling and magnetic anisotropy $\chi^{\perp}>\chi^{\parallel}$. 

On the other hand, the SC phase diagram in the presence of the Zeeman SOC is qualitatively the same irrespective of the magnitude of the interlayer hopping [Figs. \ref{fig:phase_SOC} (c) and \ref{fig:phase_SOC} (d)]. 
The staggered Zeeman SOC stabilizes the $A_{2u}$ or $E_{u}$ SC states, depending on magnitude of the Zeeman SOC. 
The $A_{2u}$ SC state is stabilized in a small Zeeman SOC region, while the $E_{u}$ SC state is stabilized in a large Zeeman SOC region. 
Both SC states are indeed compatible with the selection rule. 
The leading order parameter for the $A_{2u}$ ($E_u$) SC state is $d^{z0}$ ($\{d^{z0}_{1}, d^{z0}_{2}\}$) with $f_{x(x^2-3y^2)}$-wave ($p$-wave) symmetry [Fig. \ref{fig:gap_A2u}(a)] ([Figs. \ref{fig:gap_Eu}(c) and \ref{fig:gap_Eu}(d)]). 
Note that the leading order parameter for the $E_u$ SC state changes as $\{d^{y0}_{1}, d^{x0}_{2}\}$ ($f_{x(x^2-3y^2)}$-wave) $\to$ $\{d^{z0}_{1}, d^{z0}_{2}\}$ ($p$-wave) by increasing the SOC $\alpha_2$ so as to be compatible with the selection rule. 
The stabilization of the $E_{u}$ SC state against the $A_{2u}$ SC state may be attributed to the parity-mixing effect for the intralayer pairing. 
As shown in Fig. \ref{fig:gap_A2u}, the parity mixing effect induces an $s$-wave component $d^{0z}$ in the $A_{2u}$ SC state, and it becomes comparable to the leading $f_{x(x^2-3y^2)}$-wave component $d^{z0}$ in the large Zeeman SOC region. 
Since the $s$-wave pairing is unfavorable in the presence of the Coulomb interaction, the strongly parity-mixed $A_{2u}$ SC state is overwhelmed by the $E_u$ SC state in the large Zeeman SOC region. 
The critical value $\alpha_{2}\sim0.1$ corresponds to $\alpha_{2}=20$ meV when we adopt $t=200$ meV \cite{liu2015TMD}. 
This value lies in the realistic range of TMDs. 
Note that the competition between the selection rule and magnetic anisotropy does not occur in the case of the Zeeman SOC, in contrast to the case of the Rashba SOC. 

\subsection{Topological superconductivity \label{sec:TSC}}
Finally, we discuss the topological superconductivity. 
The $\mathbb{Z}_2$ part of topological invariants for the odd-parity SC states is determined by the occupation numbers at the TRI momenta in the Brillouin zone \cite{Sato2009,Sato2010,Fu2010}. 
In our model, the number of disconnected FSs enclosing the TRI momenta ($\Gamma$ and M points) is even. 
Thus, the $\mathbb{Z}_2$ invariant for a TRI odd-parity SC state (DI\hspace{-.1em}I\hspace{-.1em}I class) is trivial. 
The SC states that belong to one-dimensional IRs do not break the time-reversal symmetry, and hence the $A_{1u}$ and $A_{2u}$ SC states are topologically trivial. 

On the other hand, the SC states classified into 2D IRs may realize spontaneous time-reversal symmetry breaking, depending on the superposition of two gap functions. 
Then, the integer topological invariant (Chern number in class D) can be a nonzero even number. 
For instance, the $E_u$ $p$-wave SC state in a large Zeeman SOC region [Figs. \ref{fig:phase_SOC}(c) and \ref{fig:phase_SOC}(d)] should be a chiral $p_x+ip_y$ paring state in order to fully gap out the FS (i.e., the order parameter is written as $\sim(\Delta^{p_x}\pm i\Delta^{p_y})\bar{\sigma}^z\tau^0$). 
This $E_u$ $p_x+ip_y$-wave pairing state is identified as a topological SC state in class D with the Chern number  $\nu_{\rm Ch}=\pm4$ (see Appendix \ref{append:Chern}). 
A similar topological SC state is proposed in monolayer TMDs \cite{Hsu2017}, while it is a parity-mixed chiral $p+d$-wave pairing state owing to the violation of the global inversion symmetry. 

In contrast, the $E_u$ $f$-wave SC states under the Rashba SOC do not break the time-reversal symmetry, and therefore, they are topologically trivial. 
In order to fully gap out the FS, indeed, the order parameter for the $E_u$ $f_{x(x^2-3y^2)}$-wave pairing state [Fig. \ref{fig:phase_SOC}(a)] should be $\sim\Delta^f(\bar{\sigma^x}\pm\bar{\sigma^y})\tau^0$, while that for the $E_u$ interlayer pairing state [Fig. \ref{fig:phase_SOC}(b)] should be $\sim(\Delta^f\tau^x\pm\Delta^s\tau^y)\bar{\sigma^z}$. 
Time-reversal symmetry is preserved in these states. 
When we assume superposition breaking the time-reversal symmetry, the non-unitary SC state gains less condensation energy, and it is unstable.

\section{\label{sec:summary} Summary and discussion}
In summary, we have studied unconventional superconductivity in a 2D locally  NCS triangular lattice, which is relevant to the crystal structure of bilayer TMDs with 2H$_b$ stacking. 
By assuming disconnected FSs and strong electron correlation, we have clarified the dominant FM spin fluctuations on the basis of the RPA. 
The significant enhancement of the FM fluctuation is assisted by the type-I\hspace{-.1em}I vHS due to a finite interlayer coupling, and hence it is a characteristic of the bilayer structure. 
The SC instability has been discussed based on the analysis of the linearized Eliashberg equation.  
The odd-parity spin-triplet superconductivity is favored by the FM fluctuation, and we found that fully gapped $f$-wave pairing state is stabilized in a wide range of the interlayer coupling and carrier density. 
Furthermore, impacts of the staggered Rashba or Zeeman antisymmetric SOC on the magnetic fluctuation and superconductivity have been elucidated. 
The magnetic anisotropy is enhanced by increasing the SOC, and a FM-like magnetic structure with in-plane spin alignment, such as in a few-layer VSe$_2$ \cite{bonilla2018strong}, is favored by either Rashba or Zeeman SOC. 
We found that the odd-parity $A_{2u}$ or $E_u$ SC states with either $f$-wave or $p$-wave gap functions are stabilized depending on magnitude of the SOC and Coulomb interaction. 
The stability of each odd-parity SC states is determined by a combination of the selection rule for locally NCS superconductors \cite{Fischer2011,Ishizuka2018}, magnetic anisotropy, and parity-mixing effect in the SC state. 
In addition, topological properties of the stable odd-parity pairing states have been studied based on the FS formula \cite{Sato2009,Sato2010,Fu2010}. 
Then, the $E_u$ $p+ip$-wave pairing state has been identified as a topological SC state in class D with the Chern number $\nu_{\rm Ch}=4$. 
This state is stabilized by a moderate Zeeman SOC realistic in TMDs. 

Our results suggest odd-parity superconductivity ubiquitous in 2H$_b$-stacked bilayer TMDs, such as bilayer MoS$_2$ in which gate-induced superconductivity is realized \cite{costanzo2016gate,zheliuk2019josephson}. 
An essential ingredient for the odd-parity superconductivity is underlying FM fluctuations induced by a strong electron correlation. 
Although dominance of the electron-phonon coupling for the superconductivity in a few-layer TMDs is proposed by some theoretical studies \cite{Ge2013,Rosner2014,Das2015}, the electron-electron interaction is also expected to affect the superconductivity owing to the $d$-orbital character of carriers in TMDs \cite{Roldan2013,Yuan2014,yuan2015triplet,Hsu2017}. 
Thus, various bilayer TMDs have a potential for hosting FM fluctuations and odd-parity superconductivity. 
This study clarifies a way to control odd-parity SC phases by SOC and carrier doping, and to realize topological superconductivity in 2D TMDs. 

Our study also shed light on a possibility of odd-parity superconductivity in a variety of 2D magnetic van der Waals materials \cite{burch2018magnetism} not only TMDs. 
In van der Waals materials, strong enhancement of spin fluctuations, which potentially leads to unconventional superconductivity, is expected owing to the 2D nature.  
In fact, ferromagnetism has been detected in atomically thin film of CrI$_3$ \cite{huang2017layer}, Cr$_2$Ge$_2$Te$_6$ \cite{gong2017discovery}, VSe$_2$ \cite{bonilla2018strong}, V$_5$Se$_8$ \cite{nakano2019intrinsic}, and MnSe$_x$ \cite{o2018room}. 
Such FM van der Waals materials may offer a platform for multiple odd-parity SC phases.

\begin{acknowledgments}
The authors are grateful to J. Ishizuka, S. Sumita, Q. Chen, and J. Ye for helpful discussions. 
This work was supported by JSPS KAKENHI (Grants No. JP15H05884, No. JP18H04225, No. JP18H05227, No. JP18H01178, and No. 20H05159). 
S. K. is supported by a JSPS research fellowship and by JSPS KAKENHI (Grant No. 19J22122). 
\end{acknowledgments}

\appendix
\section{Symmetry of superconducting states \label{append:symmetry}}
In this appendix, we study symmetry constraints for SC states. 
First, we consider transformation of the Bloch state under space group operations. 
A creation operator of a Bloch state with spin $s$ on layer $m$ is defined as 
\begin{align}
c_{\bm{k},ms}^{\dag}=\sum_{\bm{R}}c_{s}^{\dag}(\bm{R}+\bm{r}_m) e^{-i\bm{k}\cdot\bm{R}}  , \label{eq:Bloch}
\end{align}
where $\bm{R}$ represents the position for the unit cell and $\bm{r}_m$ is the relative position of the layer $m$ in a unit cell.
Using Eq. (\ref{eq:Bloch}), the creation operator is transformed by a space group operation $g=\{p|\bm{a}\}$ as follows: 
\begin{align}
&g c_{\bm{k},ms}^{\dag} g^{-1} \nonumber\\
&= \sum_{\bm{R}} g c_{s}^{\dag}(\bm{R}+\bm{r}_m) g^{-1} e^{-i\bm{k}\cdot\bm{R}} , \nonumber\\
&= \sum_{\bm{R}} e^{-i\bm{k}\cdot\bm{R}} \sum_{s'} c_{s'}^{\dag}(p(\bm{R}+\bm{r}_m)+\bm{a}) D_{s's}^{(1/2)}(p) , \label{eq:bloch_trans}
\end{align}
where $D^{(1/2)}(p)$ is a representation matrix of the point group operation $p$ in the spin space. 
By defining $\bm{R}'+\bm{r}_{pm}\equiv p(\bm{R}+\bm{r}_m)+\bm{a}$, Eq. (\ref{eq:bloch_trans}) is rewritten as
\begin{align}
&g c_{\bm{k},ms}^{\dag} g^{-1} \nonumber\\
&= \sum_{\bm{R}'} e^{-i\bm{k}\cdot[p^{-1}(\bm{R}'+\bm{r}_{pm}-p\bm{r}_m-\bm{a})]} \nonumber\\
&\quad\times \sum_{s'} c_{s'}^{\dag}(\bm{R}'+\bm{r}_{pm}) D_{s's}^{(1/2)}(p),  \nonumber\\
&= e^{ip\bm{k}\cdot\bm{a}} \sum_{m',s'} c_{p\bm{k},m's'}^{\dag} D_{m'm,\bm{k}}^{({\rm perm})}(p,\bm{k}) D_{s's}^{(1/2)}(p). \label{eq:Bloch_G}
\end{align}
Here, we introduced a representation matrix for the permutation of layers as 
\begin{equation}
D_{m'm}^{({\rm perm})}(p,\bm{k})=e^{-ip\bm{k}\cdot(\bm{r}_{pm}-p\bm{r}_m)}\delta_{m',pm}. 
\end{equation} 

We investigate the symmetry of SC states based on the pair amplitude 
\begin{equation}
F_{ms,m's'}(\bm{k}) = \average{c_{\bm{k},ms}c_{-\bm{k},m's'}} , 
\end{equation}
which satisfies the fermionic antisymmetry 
\begin{equation}
F_{ms,m's'}(\bm{k}) = -F_{m's',ms}(-\bm{k}). \label{eq:fermion_sym}
\end{equation}
From Eq. (\ref{eq:Bloch_G}), it is revealed that the pair amplitude is transformed by a space group operation $g$ as 
\begin{align}
g F_{ms,m's'}^{\Gamma}(\bm{k}) g^{-1} 
&= \sum_{\{m_j, s_j\}}F_{m_1s_1,m_2s_2}^{\Gamma}(p\bm{k}) \mathcal{D}^{\Gamma}(g) \nonumber\\
&\times\mathcal{D}_{m_1 m_2, m m'}^{({\rm perm})}(p,\bm{k}) \mathcal{D}_{s_1 s_2, s s'}^{(1/2)}(p), \label{eq:gap_Gtrans} 
\end{align}
where the representation matrices are introduced as 
\begin{align}
\mathcal{D}_{m_1 m_2, m m'}^{({\rm perm})}(p,\bm{k})
&= D_{m_1 m}^{({\rm perm})}(p,\bm{k}) D_{m_2 m'}^{({\rm perm})}(p,-\bm{k}), \\
\mathcal{D}_{s_1 s_2, s s'}^{(1/2)}(p)
&= D_{s_1 s}^{(1/2)}(p) D_{s_2 s'}^{(1/2)}(p) , 
\end{align}
and $\mathcal{D}^{\Gamma}(g)$ is the representation matrix of the $\Gamma$ IR for the gap function. 
Whereas $\mathcal{D}^{\Gamma}(g)=\pm1$ for one-dimensional IRs, $\mathcal{D}^{\Gamma}(g)$ is $2\times2$ matrix for 2D IRs. 
Equations (\ref{eq:fermion_sym}) and (\ref{eq:gap_Gtrans}) are the symmetry constraints for the SC states. 
In the main text, the linearized Eliashberg equation is solved under these symmetry constraints for each of the IRs of $D_{3d}$ point group. 

\section{Chern number for $E_u$ pairing state \label{append:Chern}}

\begin{figure}[tbp]
\centering
   \includegraphics[width=86mm,clip]{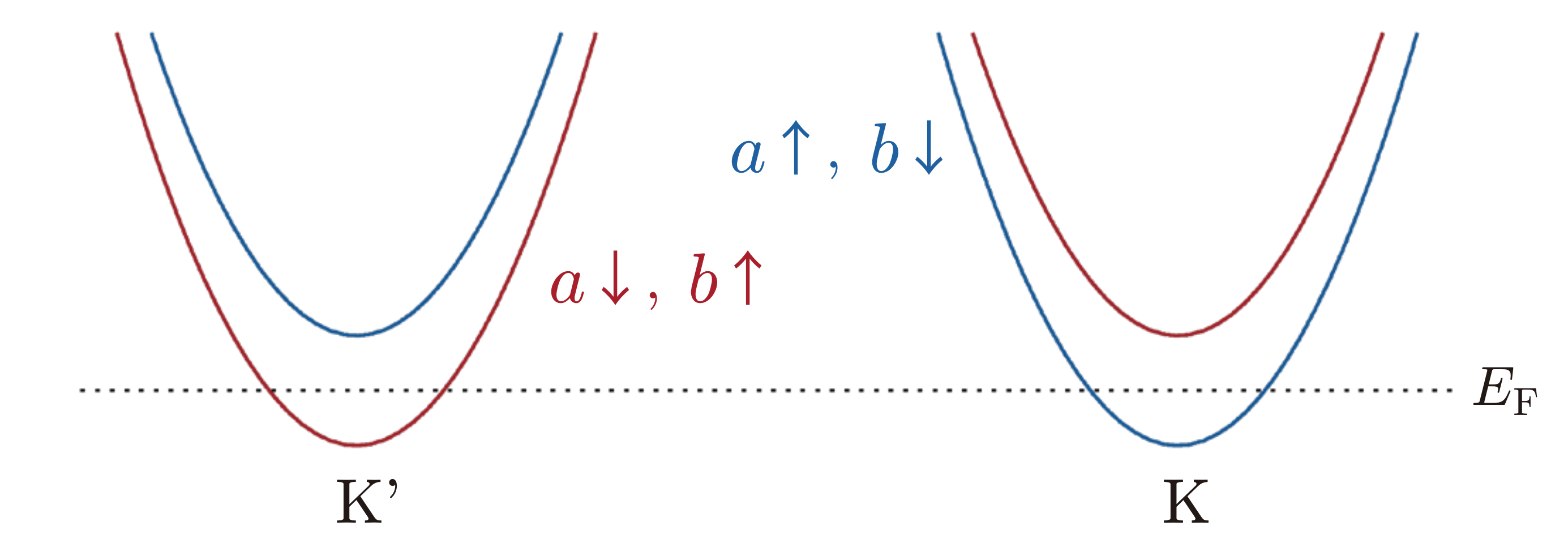}
   \caption{Schematic of the band structure near the K and K$'$ points in a large Zeeman SOC region. The Fermi energy $E_{\rm F}$ lies between the Zeeman gap. 
 \label{fig:band_Zeeman} }
\end{figure}

Here, we derive the Chern number for the $E_u$ $p_x+ip_y$-wave pairing state in a large Zeeman SOC region [Figs. \ref{fig:phase_SOC}(c) and \ref{fig:phase_SOC}(d)]. 
Figure \ref{fig:band_Zeeman} illustrates the band structure near the Fermi level under a large Zeeman SOC. 
Since the fourfold degeneracy at the K (K$'$) point is lifted by the Zeeman SOC, the band structure possesses a nearly parabolic shape around the K (K$'$) point. 
In addition, the interlayer hybridization is negligible around the K (K$'$) point owing to the threefold rotational symmetry \cite{liu2015TMD,Akashi2015,Akashi2017}. 
Then, the effective Hamiltonian for electrons near the Fermi level is derived as 
\begin{align}
\tilde{\mathcal{H}}=&
\sum_{\bm{q},m,s} \left(\tilde{\varepsilon}_{\bm{q}}-\mu \right)\psi_{\bm{q},ms}^{\dag}\psi_{\bm{q},ms} \nonumber\\
&+\frac{1}{2}\sum_{\bm{q},m,s,s'}\tilde{\Delta}_{\bm{q}}\bar{\sigma}_{ss'}^{z}\psi_{\bm{q},ms}^{\dag}\psi_{-\bm{q},ms'}^{\dag} + \mathrm{H.c.},
\end{align}
where $\tilde{\varepsilon}_{\bm{q}}=\bm{q}^2/(2m)$ is the effective kinetic energy with a parabolic dispersion, $\tilde{\Delta}_{\bm{q}}=\tilde{\Delta}^{p}(q_x+ iq_y)$ is the effective chiral $p$-wave gap function, and the annihilation operators are defined as 
$\left( \psi_{\bm{q},a\uparrow}, \,
 \psi_{\bm{q},a\downarrow}, \,
 \psi_{\bm{q},b\uparrow} , \,
 \psi_{\bm{q},b\downarrow} \right)
\equiv \left(c_{\bm{\mathrm{K}}+\bm{q},a\uparrow}, \,
c_{-\bm{\mathrm{K}}+\bm{q},a\downarrow}, \,
c_{-\bm{\mathrm{K}}+\bm{q},b\uparrow},\,
 c_{\bm{\mathrm{K}}+\bm{q},b\downarrow} \right)$. 
We assume that $\tilde{\Delta}^{p}$ is a real number. 
By using the vector operator 
\begin{align}
\hat{\Psi}_{\bm{q},m}^{\dag}=
\left(
 \psi_{\bm{q},m\uparrow}^{\dag}, \,
 \psi_{-\bm{q},m\uparrow}, \,
 \psi_{\bm{q},m\downarrow}^{\dag}, \,
 \psi_{-\bm{q},m\downarrow}, \,
\right), 
\end{align}
we obtain the matrix representation of the effective Hamiltonian as follows:
\begin{align}
\tilde{\mathcal{H}}&=\frac{1}{2}\sum_{\bm{q}}
\left(
 \hat{\Psi}_{\bm{q},a}^{\dag}, \, \hat{\Psi}_{\bm{q},b}^{\dag}
\right)
\tilde{\mathcal{H}}_{\bm{q}}
\begin{pmatrix}
 \hat{\Psi}_{\bm{q},a} \\
 \hat{\Psi}_{\bm{q},b}
\end{pmatrix}
+ \mathrm{const.} , 
\end{align}
where the Hamiltonian matrix $\tilde{\mathcal{H}}_{\bm{q}}$ is given by
\begin{align}
\tilde{\mathcal{H}}_{\bm{q}}&=\tau^0\otimes 
\begin{pmatrix}
 (\tilde{\varepsilon}_{\bm{q}}-\mu)\sigma^z 
   & \tilde{\Delta}^{p}\left(q_x\sigma^x-q_y\sigma^y\right) \\
 \tilde{\Delta}^{p}\left(q_x\sigma^x-q_y\sigma^y\right)
   & (\tilde{\varepsilon}_{\bm{q}}-\mu)\sigma^z
\end{pmatrix}.
\end{align}
Here, we carry out an unitary transformation as
\begin{align}
U \tilde{\mathcal{H}}_{\bm{q}}U^{\dag} &=\tau^0\otimes
\begin{pmatrix}
 \tilde{\mathcal{H}}^{+}_{\bm{q}}  &  0  \\
 0  & \tilde{\mathcal{H}}^{-}_{\bm{q}}
\end{pmatrix} ,
\\ 
\tilde{\mathcal{H}}^{\pm}_{\bm{q}}&=
\begin{pmatrix}
 \tilde{\varepsilon}_{\bm{q}}-\mu  &  \pm\tilde{\Delta}^{p}(q_x + iq_y)  \\
 \pm\tilde{\Delta}^{p}(q_x - iq_y)  & -\tilde{\varepsilon}_{\bm{q}}+\mu
\end{pmatrix}, \label{eq:spinless_p+ip}
\end{align}
where the unitary matrix $U$ is defined as
\begin{align}
U&=\frac{1}{\sqrt{2}}\tau^0\otimes
\begin{pmatrix}
 \sigma^0  &  \sigma^0  \\
 \sigma^0  & -\sigma^0
\end{pmatrix} .
\end{align}
Equation (\ref{eq:spinless_p+ip}) is the Bogoliubov-de Gennes Hamiltonian for the spinless chiral $p$-wave superconductivity. 
Thus, the spin-full chiral $p$-wave SC state is converted to two pairs of the spinless chiral $p$-wave SC states \cite{Hsu2017}. 
Since a spinless chiral $p$-wave SC state gives the Chern number $1$, the total Chern number of the $E_u$ chiral $p$-wave SC state is obtained as $\nu_{\rm Ch}=1\times2\times2=4$.


\nocite{*}

\providecommand{\noopsort}[1]{}\providecommand{\singleletter}[1]{#1}%

\end{document}